\documentstyle[aps,twocolumn]{revtex}

\def\be{\begin{equation}}
\def\ee{\end{equation}}

\begin{document}
\draft
%\preprint{HEP/123-qed}
%\wideabs{
\title{Comment on ``Effective Mass and g-Factor of Four Flux Quanta Composite
Fermions"}
\author{Daijiro Yoshioka}
\address{
Department of Basic Science, University of Tokyo\\
3-8-1 Komaba, Meguro-ku, Tokyo 153-8902, Japan}
\date{19 November 1998}
\maketitle
\begin{abstract}
\end{abstract}
%}
\narrowtext

%================================================================
In a recent Letter, Yeh et al.\cite{yeh} have shown beautiful experimental
results which indicate that the composite fermions with four flux quanta
($^4$CF) behave as fermions with mass and spin just like those with two
flux quanta.
They observed the collapse of the fractional quantum Hall gaps when the
following condition is satisfied with some integer $j$,
\be
g^*\mu_{\rm B}B_{\rm tot} = j \hbar \omega_{\rm c}^*,
\ee
where $g^*$ and $\omega_{\rm c}^*$ are the g-factor and the cyclotron
frequency of the $^4$CF, respectively.
They argued that level crossing and hence the collapse occur
when eq.(1) is satisfied.
However, as they have pointed out,
level crossing always occurs away from the Fermi level,
and gap at the Fermi level
remains finite in their interpretation.
Thus the reason of the collapse was left as a mystery.
Here I will show that part of the mystery is resolved by considering the
electron-hole symmetry properly.

The experiment was done around $\nu=3/4$, where $^4$CF quantum Hall effect
occurs at the electron filling factor $\nu=(3p+1)/(4p+1)$, $p= \pm 1,
\pm 2...$.
Yeh et al. considered these fillings as electron-hole symmetric states of
filling factor $\nu=p/(4p+1)$.
%Therefore in their picture lower-energy, down-spin $^4$CF filles $p$ Landau
%levels while up-spin $^4$CF Landau levels are empty in the limit of $g^* \to
%\infty$.
However, this treatment is not appropriate.
We should consider the spin freedom of the $^4$CF as coming from the actual
spin of the original electrons.
Therefore, the electron-hole symmetric state of the filling factor
$\nu=(3p+1)/(4p+1)$ is realized at $\nu = 1 + p/(4p+1)$.
%Namely, we should consider the situation where the lowest down-spin Landau
%level is fully occupied.

Now we need to establish a rule for the composite fermion transformation in
this situation.
For that purpose it is helpful to consider the case of two flux quanta
composite fermion ($^2$CF) around $\nu=1/2$.
The electron-hole symmetric state of this case is realized around $\nu=1+1/2$.
Requiring that the condition for the collapse of the quantum Hall state has
the electron-hole symmetry, we get a set of rules.
We explain these rules by applying them to the present case of the filling
factor $\nu=1 + p/(4p+1)$.
To make the discussion concrete, we consider a finite size system with $N_e$
electrons and the Landau level degeneracy $N_0$, i.e.
$\nu=1 + p/(4p+1) = N_e/N_0$.
This system is converted into that of $^4$CF by the following rules.
(1) All the electrons are changed into CF's.
(2) However, each electrononic states in the down spin Landau level gives
four flux quanta with opposite sign, thus the effective flux quanta in the
system is $N_{0,{\rm eff}} = N_0 - 4(N_e-N_0)$.
Then the effective magnetic field for the CF's is reduced from $B_{\perp}$
to $B_{\rm eff} = B_{\perp}/(4p+1)$.
The remaining rules are that (3) the maximum allowed CF filling factor for
each spin state is $|4p+1|$,\cite{comment1}
 and (4) the Zeeman splitting of these levels
are $g^*\mu_{\rm B} B_{\rm tot}$.
Therefore in the limit of infinite $g^*$ the down spin CF Landau levels are
fully occupied up to $\nu_\downarrow = |4p+1|$, and up spin levels are
occupied with filling factor $\nu_\uparrow =|p|$.\cite{comment}

Now we can investigate the condition for the collapse of the quantum Hall
state.
In the course of reducing $g^*$ from infinity, the collapse of the gap at
the Fermi level occurs when the highest occupied Landau level (HOLL) of the
down spin CF coincide with the lowest unoccupied Landau level(LULL) of the
up spin CF.
When $\nu_\uparrow = |p| + k$, and $\nu_\downarrow = |4p+1| -k$, with
integer $k$,
the energies of the down spin HOLL and up spin LULL are at
\be
E_{\uparrow,\downarrow} = \hbar\omega_c^*
(\nu_{\uparrow,\downarrow} \mp \frac{1}{2})
\mp \frac{1}{2} g^*\mu_{\rm B} B_{\rm tot},
\ee
%\be
%\hbar\omega_c^* (|p| + k + \frac{1}{2})
%+ \frac{1}{2} g^*\mu_{\rm B} B_{\rm tot},
%\ee
respectively.
Therefore the condition for the collapse is
\be
g^*\mu_{\rm B} B_{\rm tot} = (|4p+1|-|p|-2k-1) \hbar\omega_c^*
\equiv j \hbar\omega_c^*.
\ee

For negative $p$ this condition completely explains the strong collapse
of the experiment.
For example at $p=-2$ or $\nu=5/7$, the collapse is expected at $j=0$, 2,
and 4, which agrees with the experiment where among $j=1$, 2, and 3 only
$j=2$ shows the collapse.
On the other hand, for positive value of $p$ the condition above, $j=3p-2k$,
gives only weak collapse, and stronger collapses are observed at $j=3p-2k-1$.
This lack of symmetry between positive and negative values of $p$ has not
been observed for the quantum Hall states of $^2$CF around $\nu=3/2$, where
every strong collapse is associated with the collapse of the energy gap at
the Fermi level.\cite{du}
The present asymmetry may be related to the fact that the positive $p$
gives electron filling factor close to unity: Around $\nu=1$ skyrmion
excitation that mixes the up and down spin states are known to exists.

Thus I have shown that the strong collapse is related to the closing of the
energy gap at the Fermi level of the $^4$CF as long as the opposite spin
CFs are independent of each other.

I thank Dan Tsui who showed me the experimental results prior to publication
while we stayed at Aspen Center for Physics, where main part of this work
was done.

%================================================================


\begin{references}
\bibitem{comment} Notice that the large $g^*$ limit, where
$\nu$ and $1+\nu$ states should be equivalent, is recovered
correctly by these rules.
\end{references}
\end{document}